\documentclass[aps,notitlepage,onecolumn,times,pre,10pt]{revtex4-1}

\usepackage{epsfig}
\usepackage{graphicx}
\usepackage{epstopdf}
\usepackage{amssymb}
\usepackage{amsmath}
\usepackage{amsfonts}
\usepackage{mathrsfs}
\usepackage{float}
\usepackage[usenames,dvipsnames]{xcolor}
\definecolor{webgreen}{rgb}{0,0.75,0}
\definecolor{webred}{rgb}{0.75,0,0}
\definecolor{webblue}{rgb}{0,0,0.75}
\definecolor{darkblue}{rgb}{0,0,0.7}
\definecolor{dunkelgrau}{rgb}{0.8,0.8,0.8}
\definecolor{lgray}{rgb}{0.95,0.95,0.95}
\definecolor{lgreen}{rgb}{0.95,1.00,0.90}
\definecolor{lblue}{rgb}{0.9,0.95,1.00}
\definecolor{lred}{rgb}{1.00,0.90,0.80}
\definecolor{shadecolor}{rgb}{1.00,0.92,0.82}

\usepackage[breaklinks]{hyperref}
\hypersetup{colorlinks=true, linkcolor=darkblue, citecolor=darkblue, filecolor=darkblue, urlcolor=darkblue}
\usepackage[hyphenbreaks]{breakurl}

\def\dj{d\kern-0.4em\char"16\kern-0.1em}
\def\Dj{\mbox{\raise0.3ex\hbox{-}\kern-0.4em D}}

\DeclareMathOperator{\Tr}{Tr}
\DeclareMathOperator{\re}{re}
\DeclareMathOperator{\diag}{diag}
\DeclareMathOperator{\eff}{eff}
\def\dj{d\kern-0.4em\char"16\kern-0.1em}
\def\Dj{\mbox{\raise0.3ex\hbox{-}\kern-0.4em D}}
\newcommand{\mi}{\mathrm{i}}

\begin{document}

\title{Correlation between quantum entanglement and quantum coherence in the case of XY spin chains with the Dzyaloshinskii-Moriya interaction}

\author{Sonja Gombar}
\email[]{df.sonja.gombar2@student.pmf.uns.ac.rs}
\author{Petar Mali}
\author{Milan Panti\' c}
\author{Milica Pavkov-Hrvojevi\' c}
\author{Slobodan Rado\v sevi\' c}
\affiliation{Department of Physics, Faculty of Science, University of Novi Sad,
Trg Dositeja Obradovi\' ca 4, 21000 Novi Sad, Serbia}

\begin{abstract}
Recently, there has been an increased interest in studying quantum entanglement and quantum coherence. Since both of these properties are attributed to the existence of quantum superposition, it would be useful to determine whether some type of correlation between them exists. Hence, the purpose of this paper is to explore the type of the correlation in several systems with different types of anisotropy. The focus will be on the XY spin chains with the Dzyaloshinskii-Moriya interaction and the type of the mentioned bond will be explored using the quantum renormalization group method. 
\end{abstract}

 \maketitle

\section{Introduction}

Perhaps one of the most peculiar aspects of quantum mechanics, quantum information science was born due to the simple fact that it offers results achieved with higher efficiency than the ones obtained by the classical approach. In 1965 Gordon Moore made a prediction concerning the processing power for computers \cite{1}. He discovered that the number of transistors per chip would increase as its size decreased and assumed that the processing power would double every two years while the price remained constant. The main problem with this statement is the fact that transistors cannot shrink infinitely and the chip performance has noticeably slowed down recently. Some papers suggest turning to multicore chips as they tend to pick up overall performance given by single-core models by allowing some parallel work to be done \cite{2}. 

However, this transition seems less practical as the usage of classical parallel processing capability proved to be quite expensive. Therefore, some papers recommend using opportunities of quantum computing in several areas, such as in chemistry \cite{3}, instead. Moreover, both excited and ground states for $\mbox{H}_{2}$ molecule have recently been simulated on a superconducting-qubit-based processor \cite{oo}. The reason for an increased interest in this area lies in the fact that the increase in parallel-processing capability is large since classical bits 0 and 1 are replaced by quantum qubits, which can be found in quantum superposition of the analogous states. The alternative solution emerged as a spontaneous consequence of arriving at more and more strict definitions of Church-Turing thesis \cite{4} in works of Richard Feynman and David Deutsch \cite{5,6}. 

It is clear nowadays that the processing speed would increase immensely in the case of quantum computers, but that is not their only advantage. The number of solvable problems increases as well. Peter Schor and Lov Grover discussed several problems that would either be possible to solve only by using quantum computer or the amount of time necessary to obtain the answer would decrease greatly \cite{7,8}.

However, the most extraordinary benefit in the world of quantum information turns out to be quantum entanglement. It is a phenomenon present in systems containing more than one particle. Even though the particles may be separated by large distances, they  share a so-called entangled state. None of the particles may be treated independently of each other \cite{ved}. Ideas of quantum cryptography, quantum teleportation and dense coding are deeply related to the existence of the entangled states \cite{4}. In addition, a criterion for detecting the presence of entanglement in a system has been developed. It consists in checking the validity of some type of Bell's inequality (most frequently the CHSH inequality) \cite{9}.

Still, the question remains - what are the possibilities of theoretical and practical quantum information overlap? It appears that, while the theoretical aspect of quantum computation has a lot to offer, the practical solutions still cannot keep up. However, not everything is that negative. For example, quantum cryptography has already been performed for various qubit-based systems and distances \cite{10,11}. The analogous results have been achieved in the entangled higher-dimensional quantum systems (qudits), qutrits in particular \cite{12}. As a matter of fact, many properties of quantum information have been experimentally explored in recent years \cite{a,b,c,d,e,f,g,h}. 

Also, finding a system that can simulate quantum systems efficiently enough has become a problem of extreme importance. It is said that no classical computer is able to perform the simulation with high efficiency and thus the interest has turned to quantum computation \cite{4}. Many systems which are able to simulate quantum computers have been found. Some of them are optical quantum computers, which usually use polarization in order to encode quantum information, nuclear magnetic resonance, where nuclear spins are used as qubits, trapped ions and many more \cite{13}. One of the recent technologies involves quantum dots and single dopants in solid-state hosts, where large amount of qubits are easily cooled and arranged \cite{14}. However, the most common representations of quantum computers are the spin-based ones \cite{15}. Although several quantum computation systems involving a few entangled qubits have been developed, a problem of decoherence emerged and led to questioning the ways of maintaining coherence in such systems \cite{16}. Basically, one might say that decoherence comes from the system's entanglement with its environment. Therefore, during the calculation of a measure of decoherence in a certain system, one has to consider its surroundings.

The purpose of this paper is to consider a particular quantum system and observe the behaviour of two very important quantum information entities, quantum entanglement and quantum coherence, in order to explore their connection in the given system. Since spin models are used to describe a variety of different physical problems, such as the properties of magnetic compounds \cite{i1,i*,i**,i***,i****}, quantum Hall systems \cite{j,i*****}, Josephson junction arrays \cite{k}, quantum phase transitions \cite{111,222,333,444,555} etc., it is clear that quantum information is no exception. Nowadays, XY and XYZ spin chains \cite{x,y} are of significant interest, especially the ones frustrated with the Dzyaloshinskii-Moriya interaction (DM interaction)\cite{17,fip,18,19,20,211}. In this paper, the XY model with the DM interaction will be considered for systems based on three $1/2$-spins. Quantum renormalization group is a significant tool which helps one to observe the appearance of quantum phase transitions \cite{18,19,20}. However, in this article it will be used to check the validity of the conclusions obtained in the case of larger systems (containing several spin-blocks). The main goal will be to find how entanglement and coherence are related in a system detached from its environment. Therefore, no effect from the surroundings will be taken into account. 

Special attention is given to the measures of quantum entanglement and quantum coherence nowadays \cite{21,22}. However, while quite a lot is already known about entanglement, quantum coherence is still relatively unexplored. Therefore, it would be quite useful to investigate whether there exists a correlation between these two entities. Since both are a consequence of quantum superposition, only differently defined, it is not illogical to make such an assumption. Previously, the authors of \cite{37} came to the conclusion that the relative entropy of coherence and von Neumann entropy manifest the reverse behaviour. In addition, the paper outlined that the entanglement of formation of bipartite system increases if the relative entropy of coherence for one of the subsystems decreases and vice versa. However, one of the most significant papers in this area is Streltsov's article \cite{38}. It describes the possibility that the incoherent operations may generate entanglement of a state with an incoherent state, but only if the initial state was coherent. On the other hand, the aim of the present paper is to find out what is the actual correlation in the specific system in the case of different types of anisotropy when none of the operations acts on the discussed system.
 
Therefore, after a brief introduction in section 1, the purpose of section 2 is to introduce several models used in the calculations. Focus will be on three cases: the cases of a symmetric anisotropy with and without involvement of the single-ion anisotropy and the case of an asymmetric type of anisotropy. The results obtained in these cases will be presented in section 3, where the type of examined correlation will be established. The paper ends with discussion of the results in section 4. 
 
\section{Model}

The type of connection between entanglement and coherence will be explored in the case of the XY Heisenberg model with two types of anisotropy. Both of the considered Hamiltonians will include the DM interaction, as it was already proven that it gives rise to coherence \cite{19}. One may say that its presence is non-negligible in various systems, such as $\mbox{CsCuCl}_{3}$ and $\mbox{FeBO}_{3}$ \cite{23,121} and it is thus frequently theoretically explored \cite{124,125}. The interaction was first introduced by Dzyaloshinskii in 1958 \cite{24}, but its importance as a relevant effect in the real systems, which emerges as a consequence of the spin-orbital coupling, was brought up a bit later, in 1960, in Moriya's paper \cite{25}. It is also known as the antisymmetric exchange due to the fact that its origin lies in the antisymmetric part of the interaction matrix. 

The first considered case involves the Hamiltonian of an XY spin chain with a symmetric type of anisotropy and the DM interaction given by \cite{18}:
\begin{equation}
H=\frac{J}{4}\sum_{i=1}^{N}\bigg((1+\gamma)\sigma_{i}^{x}\sigma_{i+1}^{x}+(1-\gamma)\sigma_{i}^{y}\sigma_{i+1}^{y}+(-1)^{i}D(\sigma_{i}^{x}\sigma_{i+1}^{y}-\sigma_{i}^{y}\sigma_{i+1}^{x})\bigg), \label{eq:1}
\end{equation} 
where $N$ is the number of sites. However, because the transformation is unitary, the results presented in section 3 remain the same, as one can easily verify. When $\gamma=1$ and $D=0$, the model reduces to the Ising model, while for $\gamma=D=0$ it corresponds to the XX model \cite{18}. If one considers interval $0<\gamma\leq1$, models fall in the Ising universality class and for $N\to\infty$ they undergo a phase transition at the critical value of the parameter $\gamma$ \cite{26}.

However, one of the aims of the present paper will be to consider how entanglement and coherence behave when the size of the system increases. In order to arrive at such conclusion, the quantum renormalization group method will be used. Since the main goal of this method is to obtain a self-similar Hamiltonian after the necessary transformation, model has to involve $\pi$-rotation around the $x$-axis for all even sites, while the odd remain intact \cite{18}. Therefore, in the case of three-site system, the Hamiltonian \eqref{eq:1} turns into:  
\begin{equation}
H=\frac{J}{4}\sum_{i=1}^{2}\bigg( \big( 1+\gamma\big) \sigma_{i}^{x}\sigma_{i+1}^{x}-\big( 1-\gamma \big) \sigma_{i}^{y}\sigma_{i+1}^{y}+D\big(  \sigma_{i}^{x}\sigma_{i+1}^{y}+ \sigma_{i}^{y}\sigma_{i+1}^{x} \big) \bigg) \label{eq:2}
\end{equation} 
(the same Hamiltonian was previously used in \cite{18,19,26a,26b,26c}), where $J$ is the nearest neighbour coupling constant, $\gamma$ anisotropy parameter, $D$ $z$-component of the strength of the DM interaction and $\sigma_{i}^{\alpha}$ ($\alpha=x,y$) are Pauli matrices for the $i$th site defined as:
\begin{equation}
\sigma^{x}=\begin{bmatrix}
   0       & 1 \\
   1       & 0 \\
\end{bmatrix}, \hspace{2mm}
\sigma^{y}=\begin{bmatrix}
   0       & -\mi \\
   \mi       & 0 \\
\end{bmatrix}, \hspace{2mm}
\sigma^{z}=\begin{bmatrix}
   1       & 0 \\
   0       & -1 \\
\end{bmatrix}. \label{eq:3}
\end{equation}
One of the main reasons for using Hamiltonian \eqref{eq:2} is the fact that different types of coupling between neighboring spins may be controlled experimentally in qubit systems \cite{eksp1,eksp2}.

It may be seen that the shape of anisotropy is symmetric in this Hamiltonian and this will also affect the behaviour of the discussed measures. Also, the parameter $D$ in equation \eqref{eq:2} is actually the relative strength of the DM interaction because from the Hamiltonian equation one may conclude that $D$ is in fact written in terms of the coupling parameter $J$. This is the exact reason why eigenvectors turn out to be independent on $J$.

The ground state of the Hamiltonian \eqref{eq:2} is doubly degenerate and it corresponds to the eigenvalue (energy):
\begin{equation}
E_{0}=-\frac{J}{\sqrt{2}}q \label{eq:4}
\end{equation}
with $q=\sqrt{1+D^{2}+\gamma^{2}}$, and the complex eigenvectors written in the standard basis:
\begin{align}
\vert\Psi\rangle=\frac{1}{\sqrt{2}q}\sqrt{D^{2}+1}\bigg( -\frac{q}{\sqrt{2}(1+\mi D)}\vert\uparrow\uparrow\downarrow\rangle+\frac{\gamma}{1+\mi D} \vert\uparrow\downarrow\uparrow\rangle-\frac{q}{\sqrt{2}(1+\mi D)}\vert\downarrow\uparrow\uparrow\rangle+\vert\downarrow\downarrow\downarrow\rangle\bigg) \label{eq:5}
\end{align}
and:
\begin{align}
\vert \Psi'\rangle=\frac{1}{2}\bigg(-\frac{\sqrt{2}(1-\mi D)}{q}\vert\uparrow\uparrow\uparrow\rangle+\vert\uparrow\downarrow\downarrow\rangle-\frac{\sqrt{2}\gamma}{q}\vert\downarrow\uparrow\downarrow\rangle+\vert\downarrow\downarrow\uparrow\rangle\bigg), \label{eq:6}
\end{align} 
where $\vert\uparrow\rangle$ and $\vert\downarrow\rangle$ are the basis vectors of $\sigma^{z}$ Pauli matrix in the given representation. 

Also, including the single-ion anisotropy, commonly known as the magnetocrystalline anisotropy \cite{27,28}, turns out to have a trivial contribution to the previously considered case when one explores spin-$1/2$ systems. In this case, the Hamiltonian \eqref{eq:2} will have the following addition: 
\begin{equation}
H_{SI}=\sum_{i=1}^{3}\frac{JA}{4}(\sigma_{i}^{z})^{2}, \label{eq:7} 
\end{equation} 
where $A$ is the strength of the single-ion anisotropy. The previously obtained ground state energy changes to:
\begin{equation}
E_{0}=\frac{3AJ}{4}-\frac{q\vert J\vert}{\sqrt{2}} \label{eq:8}
\end{equation}
with the complex eigenvectors:
\begin{align}
\vert\Psi\rangle=\frac{1}{\sqrt{2}q}\sqrt{D^{2}+1}\bigg( -\frac{q\vert J\vert}{\sqrt{2}J(1+\mi D)}\vert\uparrow\uparrow\downarrow\rangle+\frac{\gamma}{1+\mi D} \vert\uparrow\downarrow\uparrow\rangle-\frac{q\vert J\vert}{\sqrt{2}J(1+\mi D)}\vert\downarrow\uparrow\uparrow\rangle+\vert\downarrow\downarrow\downarrow\rangle\bigg) \label{eq:9}
\end{align}
and:
\begin{align}
\vert \Psi'\rangle=\frac{1}{2}\bigg(-\frac{\sqrt{2}J(1-\mi D)}{q\vert J\vert}\vert\uparrow\uparrow\uparrow\rangle+\vert\uparrow\downarrow\downarrow\rangle-\frac{\sqrt{2}J\gamma}{q\vert J\vert}\vert\downarrow\uparrow\downarrow\rangle+\vert\downarrow\downarrow\uparrow\rangle\bigg). \label{eq:10}
\end{align} 
Therefore, one can already assume that the single-ion anisotropy will not affect the considered entities. This was already intuitively expected as this sort of anisotropy is trivial in the case of spin-$1/2$, when it contributes to the Hamiltonian \eqref{eq:2} only as an addition in the identity matrix multiplied by a constant. Therefore, a conclusion can be made: the single-ion anisotropy does not affect neither entanglement nor coherence when the considered spins are equal to $1/2$. 

The second case will involve an asymmetric type of anisotropy. The Hamiltonian in this case is given by:
\begin{equation}
H=\frac{\tilde{J}}{4}\sum_{i=1}^{2}\bigg(  \sigma_{i}^{x}\sigma_{i+1}^{x}+M \sigma_{i}^{y}\sigma_{i+1}^{y}+\tilde{D}\big(  \sigma_{i}^{x}\sigma_{i+1}^{y}+ \sigma_{i}^{y}\sigma_{i+1}^{x} \big) \bigg), \label{eq:11} 
\end{equation}
where $\tilde{J}$ is once again the nearest neighbour coupling constant, $M$ anisotropy parameter and $\tilde{D}$ is the $z$-component of the strength of the DM interaction. It is easy to show that the Hamiltonian \eqref{eq:11} turns into \eqref{eq:2} by the simple transformation:
\begin{equation}
\tilde{J}=(1+\gamma)J, \hspace{3mm} \tilde{D}=\frac{D}{1+\gamma}, \hspace{3mm} M=\frac{\gamma-1}{\gamma+1} \hspace{1mm}. \label{eq:12}
\end{equation}
However, it can already be noticed that the problem is that this sort of transformation is not defined in the case $\gamma=-1$ and thus it would not be reasonable to call this model more general than the previous one, but rather a model with a different kind of anisotropy. In this case, the obtained ground state corresponds to the energy:
\begin{equation}
E_{0}=-\frac{1}{2}\tilde{J}m, \hspace{3mm} m=\sqrt{1+2\tilde{D}^{2}+M^{2}} \label{eq:13}
\end{equation} 
and the complex eigenstates:
\begin{align}
\vert\Psi\rangle=\frac{\sqrt{4\tilde{D}^{2}+(-1+M)^{2}}}{2m}\bigg(&\frac{\mi m}{2\tilde{D}+\mi(-1+M)}\vert\uparrow\uparrow\downarrow\rangle-\frac{\mi(1+M)}{2\tilde{D}+\mi(-1+M)}\vert\uparrow\downarrow\uparrow\rangle \nonumber \\ &+\frac{\mi m}{2\tilde{D}+\mi(-1+M)}\vert\downarrow\uparrow\uparrow\rangle+\vert\downarrow\downarrow\downarrow\rangle\bigg) \label{eq:14}
\end{align}
and:
\begin{equation}
\vert\Psi'\rangle=\frac{1}{2}\bigg(\frac{-1+2\mi \tilde{D}+M}{m}\vert\uparrow\uparrow\uparrow\rangle+\vert\uparrow\downarrow\downarrow\rangle-\frac{1+M}{m}\vert\downarrow\uparrow\downarrow\rangle+\vert\downarrow\downarrow\uparrow\rangle\bigg), \label{eq:15}
\end{equation}
which are considerably different than functions \eqref{eq:5} and \eqref{eq:6} and thus a difference in the behaviour of the quantum information entities is expected.
\par It turns out that there are a few measures that satisfy conditions every valid measure of entanglement needs to satisfy \cite{21}. The most frequent are the entanglement of formation, the entanglement of distillation and the relative entropy of entanglement. However, adapting these measures to particular systems can be a serious problem. Although some measures for multiparticle entanglement have been found \cite{31,32}, a measure originally found by Wooters, the concurrence for bipartite systems will be used \cite{33}:
\begin{equation}
C(\rho)=\max\lbrace 0,\lambda_{1}-\lambda_{2}-\lambda_{3}-\lambda_{4}\rbrace, \label{eq:16}
\end{equation}
where $\lambda_{i}$ $(i=1...4)$ are the square roots of the eigenvalues of the matrix $R=\rho\tilde{\rho}$ in descending order and the matrix $\tilde{\rho}$:
\begin{equation}
\tilde{\rho}=\big(\sigma^{y}\otimes\sigma^{y}\big)\rho^{*}\big(\sigma^{y}\otimes\sigma^{y}\big), \label{eq:17}
\end{equation}
where $\rho^{*}$ is the complex conjugate of a density matrix $\rho$. Density matrix has already reached the level of one of the crucial methods in quantum mechanics and it is widely used in modern atomic physics, for describing scattering or laser physics, statistical physics, etc. \cite{l}. In this case the density matrix obtained from eigenstates of the considered Hamiltonians needs to be reduced to its bipartite form.

On the other hand, quantum coherence is a relatively new feature in the world of quantum information. Early study was released in 2006 by Aberg, who gave an approach to quantifying superposition of orthogonal quantum states \cite{34}, while the maximum was reached in 2014 in the paper dedicated to finding several conditions every valid measure of coherence needs to fulfill \cite{35}.
Several measures satisfy these conditions and the most common are the $l_{1}$ norm and the relative entropy of coherence. The latter one will be used in this paper and it is defined as:
\begin{equation}
C_{\re}(\rho)=S(\rho_{\diag})-S(\rho), \label{eq:18}
\end{equation}
where $S(\rho)$ is the von Neumann entropy given by:
\begin{equation}
S(\rho)=-\Tr \rho\log_{2} \rho\hspace{1mm}. \label{eq:19}
\end{equation}
Here $\rho_{\diag}$ represents a diagonal matrix such that all off-diagonal elements in $\rho$ are replaced with zeros. Since in this case pure states are considered (for which $S(\rho)=0$ since $\log_{2}(1)=0$), only the diagonal part remains in \eqref{eq:18}.

\section{Results}

In this section, behaviour of quantum entanglement and quantum coherence is inspected in two mentioned cases and the results are shown in Figures $1-4$ as a dependence of two quantities on the anisotropy and the DM interaction parameters.
 
\subsection{XY model with the DM interaction and the symmetric anisotropy}

According to \eqref{eq:16} one has to obtain a density matrix in order to calculate the concurrence. Since the considerations do not involve the influence of temperature, system is assumed to be in its ground state. Therefore, the density matrix of the ground state is defined as:
\begin{equation}
\rho=\vert\Psi\rangle\langle\Psi\vert=\begin{bmatrix}
0 &0 &0 &0 &0 &0 &0 &0 \\ 
0 &\frac{1}{4} &-\frac{\gamma}{2\sqrt{2}q} &0 &\frac{1}{4} &0 &0 &\frac{-1+\mi D}{2\sqrt{2}q} \\
0 &-\frac{\gamma}{2\sqrt{2}q} &\frac{\gamma^{2}}{2q^{2}} &0 &-\frac{\gamma}{2\sqrt{2}q} &0 &0 &\frac{\gamma(1-\mi D)}{2q^{2}} \\
0 &0 &0 &0 &0 &0 &0 &0 \\
0 &\frac{1}{4} &-\frac{\gamma}{2\sqrt{2}q} &0 &\frac{1}{4} &0 &0 &\frac{-1+\mi D}{2\sqrt{2}q} \\
0 &0 &0 &0 &0 &0 &0 &0 \\ 
0 &0 &0 &0 &0 &0 &0 &0 \\
0 &-\frac{1+\mi D}{2\sqrt{2}q} &\frac{\gamma(1+\mi D)}{2q^{2}} &0 &-\frac{1+\mi D}{2\sqrt{2}q} &0 &0 &\frac{1+D^{2}}{2q^{2}}
\end{bmatrix}\hspace{1mm}. \label{eq:20}
\end{equation}
Since it can be easily shown that the obtained results for two entities remain the same if one considers the other eigenvector of the ground state \eqref{eq:6}, the relevant eigenstate will be taken as \eqref{eq:5} further on.  
\par
\begin{figure}[H]
\includegraphics[scale=0.85]{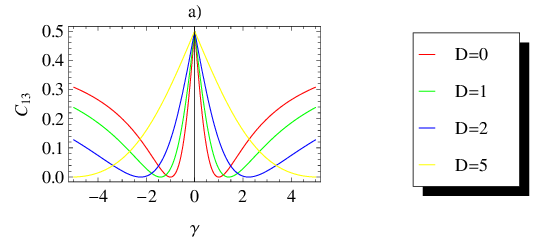} 
\includegraphics[scale=0.85]{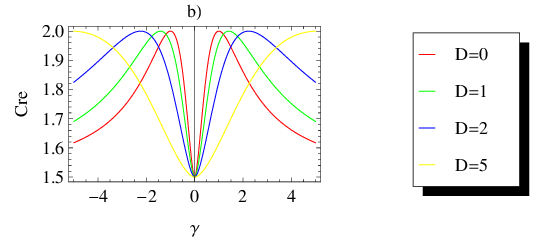} \\
\centering
\caption{Dependence of the concurrence (a) and the relative entropy of coherence (b) on the anisotropy parameter $\gamma$ and the parameter of the DM interaction in the case of three sites.}
\label{fig:f1}
\end{figure}

 However, as it was already mentioned, the relevant measure of entanglement will be the concurrence for two-qubit  systems and thus the reduced density matrix will be determined. There are two options for obtaining this matrix: calculating the concurrence between sites 1 and 3 by summing over the degrees of freedom of the second site and calculating the concurrence between the middle site and any other while summing over the degrees of freedom of the remaining site \cite{18}. The first choice will be taken, but the result is the same:
\begin{equation}
\rho^{13}=\Tr_{2}\rho=\begin{bmatrix}
\frac{\gamma^{2}}{2q^{2}} & 0 & 0 & \frac{\gamma(1-\mi D)}{2q^{2}} \\ 0 & \frac{1}{4} & \frac{1}{4} & 0 \\ 0 & \frac{1}{4} & \frac{1}{4} & 0 \\ \frac{\gamma(1+\mi D)}{2q^{2}} & 0 & 0 & \frac{1+D^{2}}{2q^{2}}
\end{bmatrix} . \label{eq:21}
\end{equation} 
Now the concurrence can easily be calculated by \eqref{eq:16} and the result is \cite{18}:
\begin{equation}
C_{13}=\sqrt{\frac{1}{4}}-\sqrt{\frac{(1+D^{2})\gamma^{2}}{q^{4}}} \hspace{1mm}. \label{eq:22}
\end{equation}
Dependence of the concurrence on the anisotropy parameter $\gamma$ and the DM parameter $D$ is shown in Figure \ref{fig:f1} a). One can observe that for small values of the anisotropy parameter the concurrence increases with the increase in the DM interaction, while for small values of the DM interaction and the larger parameter of anisotropy the concurrence decreases with the increase in the parameter $D$. This leads to the conclusion that the DM interaction tends to suppress the concurrence, while the anisotropy restores it. 

The relative entropy of coherence is calculated according to \eqref{eq:18} and the obtained result is \cite{20}:
\begin{align}
C_{\re}&=1-\frac{\gamma^{2}}{2q^{2}}\log_{2}\bigg(\frac{\gamma^{2}}{2q^{2}}\bigg)-\frac{1+D^{2}}{2q^{2}}\log_{2}\bigg( \frac{1+D^{2}}{2q^{2}}\bigg) \hspace{1mm}. \label{eq:23} 
\end{align}

Dependence of the relative entropy of coherence on the anisotropy parameter $\gamma$ and the DM parameter $D$ is shown in Figure \ref{fig:f1} b). It can be seen that this sort of behaviour is basically reverse with regard to the previously described behaviour of the concurrence.
\par According to Figure \ref{fig:f1}, it is obvious that the concurrence and the relative entropy of coherence manifest the reverse type of behaviour. The positions of the maxima of one entity correspond to the positions of the minima of the other. Therefore, one can say that the same processes that give rise to one of the properties result in the decrease of the other. However, one can notice that, while the minimum of the concurrence is positioned at $C_{13}=0$, there exists some residual relative entropy of coherence at its minimum $C_{\re}=1.5$. Nonetheless, the behaviour remains reverse. Besides, the behaviour of both entities is symmetric about the $y$-axes, which is the consequence of the symmetric choice of anisotropy $\gamma$.

Now it would be interesting to see if this conclusion stands if the system gets larger.

\subsubsection{Generalization to the larger systems}

\begin{figure}[H]
\includegraphics[scale=0.85]{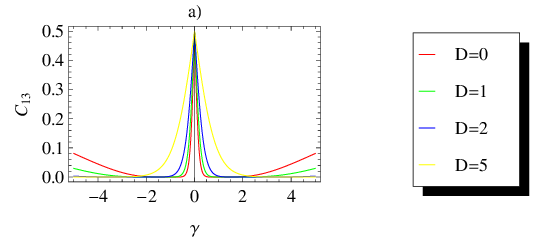} 
\includegraphics[scale=0.85]{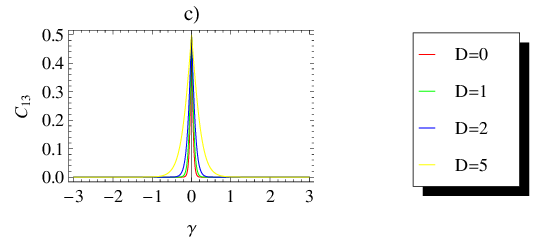} \\
\includegraphics[scale=0.85]{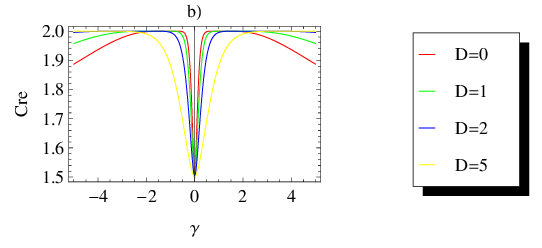}
\includegraphics[scale=0.85]{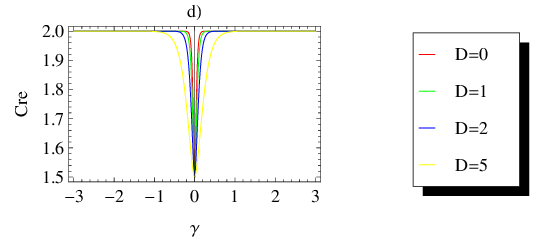} \\
\centering
\caption{Dependence of the concurrence and the relative entropy of coherence on the anisotropy parameter $\gamma$ and the parameter of the DM interaction $D$ in the case of 9 (a, b) and 27 sites (c, d).}
\label{fig:f2}
\end{figure}

In this section, a generalization to the larger systems will be made with the help of the quantum renormalization group method (QRG method). This method is essential in spin-chain problems due to the fact that a problem is reduced to considering lower dimensional systems in more convenient space \cite{19}. In the paper, the block-matrix approach designed by Kadanoff will be used \cite{36}. It consists in separating Hamiltonian to block and inter-block Hamiltonian and acquiring the basis of new, renormalized space from the ground state of block Hamiltonian. The effective Hamiltonian is acquired as \cite{18}:
\begin{equation}
H_{\eff}=\frac{J^{'}}{4}\sum_{j=1}^{N/3}\bigg((1+\gamma^{'})\sigma_{j}^{x}\sigma_{j+1}^{x}-(1-\gamma^{'})\sigma_{j}^{y}\sigma_{j+1}^{y}+D^{'}(\sigma_{j}^{x}\sigma_{j+1}^{y}+\sigma_{j}^{y}\sigma_{j+1}^{x})\bigg) \label{eq:24} 
\end{equation}
with the renormalized parameters:
\begin{equation}
J^{'}=\frac{1+D^{2}+3\gamma^{2}}{2q^{2}}J, \hspace{3mm} \gamma^{'}=\frac{3\gamma+3\gamma D^{2}+\gamma^{3}}{1+D^{2}+3\gamma^{2}}, \hspace{3mm} D'=-D \hspace{1mm}. \label{eq:25} 
\end{equation}  
Dependence of the concurrence and the relative entropy of coherence on the parameters $\gamma$ and $D$ in the case of 9 and 27 sites is shown in Figure \ref{fig:f2}.

Once again, the reverse behaviour of two quantum information properties is apparent. The minima of one entity match the maxima of the other and vice versa. Also, it can be noticed that the concurrence has sharper and sharper maxima as the size of the system increases, while the relative entropy of coherence has sharper and sharper minima. Therefore, when the size of the system $N\to\infty$, coherence has minimal value only at $\gamma=0$ and its maximal value everywhere else, while the concurrence will have non-zero value only there and zero value everywhere else. This implies that the quantum phase transition is present at $\gamma=0$ and these measures of quantum information can reveal it. However, the parameter $D$ has no effect on this behaviour and thus manipulating this parameter will not lead the system to a quantum phase transition \cite{18}.

\subsection{XY model with the DM interaction and the asymmetric anisotropy}

\begin{figure}[H]
\includegraphics[scale=0.85]{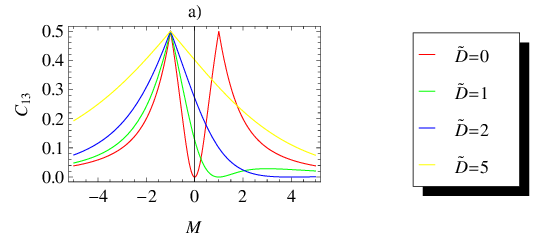} 
\includegraphics[scale=0.85]{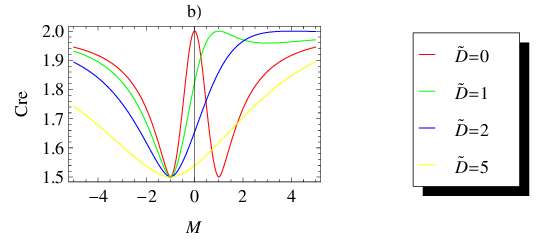} \\
\centering
\caption{Dependence of the concurrence (a) and the relative entropy of coherence (b) on the anisotropy parameter $M$ and the parameter of the DM interaction $\tilde{D}$ in the case of three sites.}
\label{fig:f3}
\end{figure}

In this case, the relevant density matrix of the ground state is:
\begin{equation}
\rho=\begin{bmatrix}
&0 &0 &0 &0 &0 &0 &0 &0 \\ &0 &\frac{1}{4} &-\frac{1+M}{4m} &0 &\frac{1}{4} &0 &0 &\frac{2\mi \tilde{D}+(-1+M)}{4m} \\ &0 &-\frac{1+M}{4m} &\frac{(1+M)^{2}}{4m^{2}} &0 &-\frac{1+M}{4m} &0 &0 &\frac{-2\mi \tilde{D}(1+M)-(M^{2}-1)}{4m^{2}} \\ &0 &0 &0 &0 &0 &0 &0 &0 \\ &0 &\frac{1}{4} &-\frac{1+M}{4m} &0 &\frac{1}{4} &0 &0 &\frac{2\mi \tilde{D}+(-1+M)}{4m} \\ &0 &0 &0 &0 &0 &0 &0 &0 \\ &0 &0 &0 &0 &0 &0 &0 &0 \\ &0 &\frac{-2\mi \tilde{D}+(-1+M)}{4m} &\frac{2\mi \tilde{D}(1+M)-(M^{2}-1)}{4m^{2}} &0 &\frac{-2\mi \tilde{D}+(-1+M)}{4m} &0 &0 &\frac{4\tilde{D}^{2}+(-1+M)^{2}}{4m^{2}} \\ 
\end{bmatrix} \label{eq:26}
\end{equation} 
and the corresponding reduced density matrix has the following form:
\begin{equation}
\rho^{13}=\begin{bmatrix}
&\frac{(1+M)^{2}}{4m^{2}} &0 &0 &\frac{-2\mi \tilde{D}(1+M)-(M^{2}-1)}{4m^{2}} \\ &0 &\frac{1}{4} &\frac{1}{4} &0 \\ &0 &\frac{1}{4} &\frac{1}{4} &0 \\ &\frac{2\mi \tilde{D}(1+M)-(M^{2}-1)}{4m^{2}} &0 &0 &\frac{4\tilde{D}^{2}+(-1+M)^{2}}{4m^{2}}
\end{bmatrix} \hspace{1mm}. \label{eq:27}
\end{equation}
Therefore, the concurrence is given by:
\begin{equation}
C_{13}=\sqrt{\frac{1}{4}}-\sqrt{\frac{(1+M)^{2}(1+4\tilde{D}^{2}-2M+M^{2})}{4m^{4}}} \label{eq:28}
\end{equation} 
and its dependence on the parameters $M$ and $\tilde{D}$ is given in Figure \ref{fig:f3} a).

Analogously, the relative entropy of coherence is obtained as:
\begin{equation}
C_{\re}=1-\frac{(1+M)^{2}}{4m^{2}}\log_{2}\bigg(\frac{(1+M)^{2}}{4m^{2}}\bigg)-\frac{4\tilde{D}^{2}+(-1+M)^{2}}{4m^{2}}\log_{2}\bigg(\frac{4\tilde{D}^{2}+(-1+M)^{2}}{4m^{2}}\bigg) \label{eq:29}
\end{equation}
and its dependence on the parameters $M$ and $\tilde{D}$ is given in Figure \ref{fig:f3} b).

The reverse behaviour can be observed once again, in the case of another type of anisotropy. The maxima of one property match the minima of the other. Therefore, one can say that the change of the anisotropy does not result in the change of the previously described correlation between the quantum information properties. The residual coherence is present at its minima again. However, this time, although the type of connection is the same as the last time, the concurrence peaks are sharper than the ones the relative entropy of coherence has. Also, $y$-axes are not the axes of symmetry this time, except in the case $\tilde{D}=0$. This is the consequence of the choice of anisotropy.

\subsubsection{Generalization to the larger systems}

\begin{figure}[H]
\includegraphics[scale=0.85]{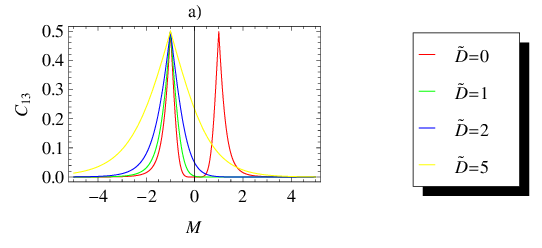} \includegraphics[scale=0.85]{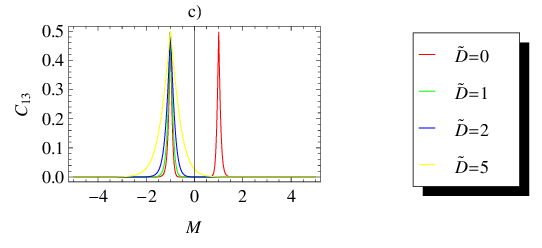}\\
\includegraphics[scale=0.85]{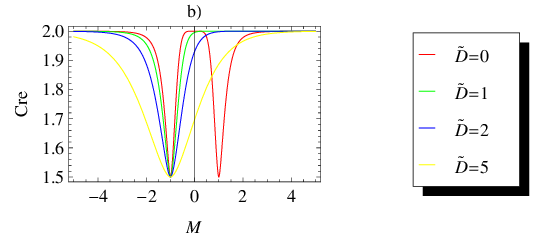} \includegraphics[scale=0.85]{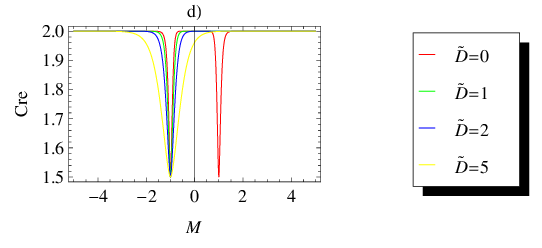}\\
\centering
\caption{Dependence of the concurrence and the relative entropy of coherence on the anisotropy parameter $M$ and the parameter of the DM interaction in the case of 9 (a, b) and 27 sites (c, d).}
\label{fig:f4}
\end{figure}

As a method to explore the correlation between entanglement and coherence in the case of the larger systems, the QRG method will be used once again. The effective Hamiltonian is given by:
\begin{equation}
H_{\eff}=\frac{\tilde{J}'}{4}\sum_{i=1}^{N/3}\bigg( \sigma_{i}^{x}\sigma_{i+1}^{x}+M' \sigma_{i}^{y}\sigma_{i+1}^{y}+\tilde{D}'\big(  \sigma_{i}^{x}\sigma_{i+1}^{y}+ \sigma_{i}^{y}\sigma_{i+1}^{x} \big) \bigg) , \label{eq:30}
\end{equation}
where the renormalized parameters are:
\begin{equation}
\tilde{J}'=\tilde{J}\frac{1+\tilde{D}^{2}(2+M)}{m^{2}}, \hspace{3mm} M'=\frac{\tilde{D}^{2}+2\tilde{D}^{2}M+M^{3}}{1+2\tilde{D}^{2}+\tilde{D}^{2}M}, \hspace{3mm} \tilde{D}'=-\tilde{D}\frac{1+\tilde{D}^{2}+M+M^{2}}{1+2\tilde{D}^{2}+\tilde{D}^{2}M} \hspace{1mm}. \label{eq:31} 
\end{equation}
Dependence of the concurrence and the relative entropy of coherence on the parameters $M$ and $D$ in the case of 9 and 27 sites is shown in Figure \ref{fig:f4}.

The reverse behaviour is obvious in the case of larger sizes of the system with the asymmetric anisotropy, too. Therefore, the size of the system has no effect on the correlation between entanglement and coherence. Also, the sharpening of the peaks is evident as the system becomes larger. This implies that the anisotropy parameter $M$ may cause the quantum phase transition, but this time for the critical value $M=-1$ due to the choice of anisotropy. Actually, another critical point may be detected at $M=+1$ only in the case $\tilde{D}=0$ and thus a quantum phase transition may be induced by the DM interaction when the chosen anisotropy is asymmetric.
\par On the other hand, the peak corresponding to $M=+1$ and $\tilde{D}=0$ does not appear in the case of the symmetric choice of anisotropy. If one chooses $\gamma=D=\tilde{D}=0$, the Hamiltonians \eqref{eq:1} and \eqref{eq:2} have the same form as \eqref{eq:11} for $M=+1$ and $M=-1$, respectively. The conducted $\pi$-rotation, which is the connecting link between \eqref{eq:1} and \eqref{eq:2}, allows one to transfer from one choice of the parameter $M$ to another (from +1 to -1 and vice versa). The model with the asymmetric choice of anisotropy covers both \eqref{eq:1} and \eqref{eq:2} for $\gamma=0$. Therefore, only the critical peak corresponding to the choice $\gamma=0$ may be detected in Figures \ref{fig:f1} and \ref{fig:f2}, while two critical peaks may be seen in Figures \ref{fig:f3} and \ref{fig:f4} at $M=-1$ and $M=+1$ when $\tilde{D}=0$.    

\section{Conclusion}

The authors of \cite{37,38,39} have already discussed the type of the correlation that exists between quantum entanglement and quantum coherence in several situations. In this particular paper a connection between these two entities is found in the case of XY spin-$1/2$ chains with the DM interaction. The purpose was to establish the nature of these correlations. Therefore, the behaviour of both measures was explored by changing the value of the DM parameter, the choice and value of anisotropy and the size of the system.   

In the case of the symmetric choice of anisotropy $\gamma$, the type of connection between the pairwise concurrence and the relative entropy of coherence is established in the present paper. This has actually proven to be a really simple type of correlation, where the properties manifest reverse behaviour. In fact, it was determined that varying values of both DM parameter and anisotropy parameter does not affect this reverse type of correlation. In this paper it was also confirmed that this type of connection remains the same when the size of the system changes, i.e. the correlation does not change as the number of the QRG iterations increases. Therefore, this statement remains valid even for the infinitely large systems.  

The next step was to consider if this reverse behaviour is persistent when one changes the choice of anisotropy and thus the asymmetric type of anisotropy was taken into account. It turned out that a different choice of anisotropy does not change the previous type of connection between the quantum information quantities and the behaviours are once again reverse.

As a conclusion, the pairwise concurrence and the relative entropy of coherence exhibit reverse behaviour when the value of the DM parameter, the choice and value of anisotropy and the size of the system are varied. Since the importance of maintaining coherence in the system has already been outlined, it is easy to see how this conclusion may be significant for the development of quantum information theory. The maximum of coherence in the system is reached when the minimum of pairwise entanglement is obtained. Therefore, in this particular system manipulating the parameters can lead the system into the state of minimal entanglement and maximal coherence. However, since both of the entities are important for quantum information, these extreme values might not be the best possible choice.   

Both quantum coherence and quantum entanglement were explored as the quantities that can indicate the existence of the quantum phase transitions (QPT) in the system with the symmetric type of anisotropy \cite{18,19}. The authors of \cite{18,19} established the appearance of the QPT for the critical value of the anisotropy parameter $\gamma_{C}=0$. Since the extreme values of both coherence and concurrence at this point remain the same regardless of the DM interaction parameter $D$, the system cannot exhibit a QPT induced by the DM interaction. However, the present paper shows that this is not the case when the asymmetric type of anisotropy is considered. First of all, the choice of anisotropy displaces the critical point to the position $M_{C}=-1$. Moreover, this is not the only critical point since the peak present at $M=+1$ for $\tilde{D}=0$ sharpens as the number of the QRG steps increases. Therefore, another critical point may be detected at $M_{C}=+1$, $D_{C}=0$. This implies that although the DM interaction cannot influence the QPT when the anisotropy is symmetric, it is most certainly able to when the asymmetric anisotropy is chosen in the case of XY spin-$1/2$ chains. 

\section*{Acknowledgment}
This work was supported by the Serbian Ministry of Education
and Science: Grant No 171009. and by the Provincial Secretariat for
High Education and Scientific Research of Vojvodina (Project
No. APV 114-451-2201).

\end{document}